\documentclass[dvips]{article}

\usepackage{icrctc07}

\title{H.E.S.S. Galactic Plane Survey unveils a Milagro Hotspot}
\shorttitle{H.E.S.S. Galactic Plane Survey unveils a Milagro Hotspot}

\authors{A. Djannati-Atai$^{1}$, E. O\~na-Wilhelmi$^{1}$,
M. Renaud$^{2}$ \& S.~Hoppe$^{2}$ for the H.E.S.S. Collaboration$^{3}$} 
\shortauthors{Djannati-Atai et al.}
\afiliations{$^1$APC, 11 Place Marcelin Berthelot, F-75231 Paris Cedex
05, France\\ $^2$Max-Plank-Institute fur Kernphysik, P.O. Box 103980,
D 69029 Heidelberg, Germany \\$^3$ \rm{ \texttt{www.mpi-hd.mpg.de/HESS}}}
\email{djannati@apc.univ-paris7.fr, emma@apc.univ-paris7.fr}

\abstract{
We report here on a new VHE source, HESS~J1908+063, disovered during the extended
H.E.S.S. survey of the Galactic plane and which coincides with
the recently reported MILAGRO unidentified source MGRO~J1908+06.
The position, extension and spectrum measurements of the HESS
source are presented and compared to those of MGRO~J1908+06. Possible
counterparts at other wavelenghts are discussed.
For the first time one of the low-lattitude MILAGRO sources is
confirmed.

}

\begin{document}
\maketitle
\section{Introduction}
H.E.S.S. observations of the inner Galactic plane in the [$270^{\circ}$, $30^{\circ}$]
longitude range have revealed more than two dozens of  new VHE
sources, consisting of shell-type SNRs, pulsar wind nebulae,
X-ray binary systems, a putative young star cluster, etc, and yet
unidentified objects (see e.g. \cite{HESSScanII} and  \cite{HESSSurveyICRC07} in
these proceedings for a summary). 

The extended  H.E.S.S. survey in the
[$30^{\circ}$-$60^{\circ}$] longitude range performed between 2005 and
2007 overlaps with regions covered by the MILAGRO sky survey at longitudes
greater than $30^{\circ}$.
The latter experiment has recently reported \cite{MILAGRO} three
low-latitude sources including,  MGRO~J1908+06,
detected after seven years of operation (2358 days of data) at
8.3$\sigma$ (pre-trials) confidence level. 
MGRO~J1908+06, of which the extension remains unknown but bounded to
a maximum diameter of 2.6$^{\circ}$, is located near the 
galactic longitude $\sim 40^{\circ}$ and hence is covered by the
H.E.S.S. galactic plane survey. 

A new H.E.S.S. source, HESS~J1908+063, which coincides
with MGRO~J1908+06, is presented here. Its position, size and spectrum
are measured and compared to the MILAGRO source. 
Possible counterparts at other wavelengths are discussed in the light
of the H.E.S.S. measurements.

\section{Observations, Analysis \& Results}
\label{results}
Results presented in this section should be
considered as preliminary.

Observations around HESS~J1908+063 were first performed during June 2005
and then from May to September 2006
as part of the extension of the Galactic plane survey in the range of
galactic longitude and latitude of 30$^{\circ}$ $<$~l~$<$60$^{\circ}$ and
$-3^{\circ}$~$<$~b~$<$3$^{\circ}$, respectively. Followup observations
were made during May and June 2007.  
In the available data-set the source is offset from the field of
view center, at different angular distances with an
average offset of 1.4$^{\circ}$. Observations for which the
source is offset by more than 2.5$^{\circ}$ were not considered for the
analysis. The total dead-time corrected and quality selected data-set
amounts to 14.9~hours with the zenith angle ranging from 30
to 46$^{\circ}$ and with a mean energy threshold  of $\sim$300~GeV. 

\begin{figure}[!t]
\begin{center}
\includegraphics[width=0.48\textwidth,angle=0,clip]{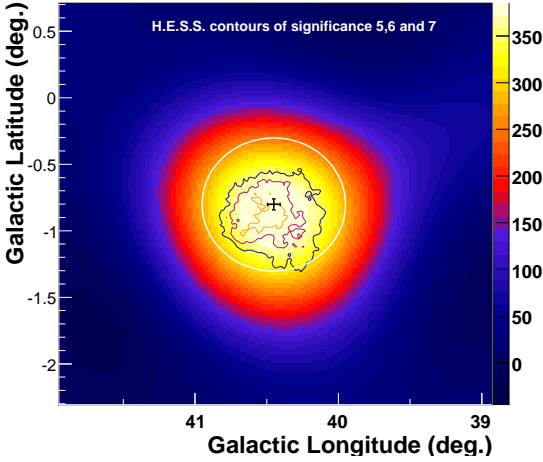}
\end{center}
\caption{ Smoothed excess map
($\sigma=0.5^{\circ}$) of the 1.5$^{\circ}\times1.5^{\circ}$ field of
view around the position of HESS~J1908+063. The contours show
the pre-trials significance levels for 5, 6 and 7$\sigma$, while the white
    circle shows the  $0.5^{\circ}$ integration radius used for the
    spectrum derivation. }
\label{skymap}
\end{figure}

After calibration, the standard H.E.S.S. event reconstruction scheme
was applied to the data \cite{HESSCrab}. In order to
reject the background of cosmic-ray showers, $\gamma$-ray like
events were selected using cuts on image shape scaled with their
expected values obtained from Monte Carlo simulations. As described in
\cite{HESSKooka}, two different sets of cuts, depending on the image
size, were applied. Cuts optimized for a hard spectrum and a weak
source with a rather tight cut on the image size of 200 p.e. (photoelectrons), 
which achieve a maximum signal-to-noise ratio, were applied to
study the morphology of the source, while for the spectral analysis,
the image size cut is loosened to 80 p.e. in order to cover the
maximum energy range. The background estimation
(described in \cite{HESSBack}) for each position in the
two-dimensional sky map is computed from a ring with an (apriori)
increased radius of $1.0^{\circ}$, as compared to the standard radius
of $0.5^{\circ}$, in order to deal with the large source diameter.
This radius yields four times a larger area for the background estimation
than the considered on-source region. Also events coming from known sources were
excluded to avoid contamination of the background. For the spectrum analysis, the
background is evaluated from positions in the field of view with the
same radius and same offset from the pointing direction as the source region.

Fig.~\ref{skymap} shows the Gaussian-smoothed excess map for a size
cut on the images above 200 p.e. The 
colored contours indicate the H.E.S.S. pre-trials significance contour
levels for 5,6 and 7 $\sigma$. HESS~J1908+063 was discovered first as a
hot-spot within the standard survey analysis scheme \cite{HESSScanII} and was
subsequently confirmed at 7.7 $\sigma$ (pre-trials). 
A conservative estimate of the trials yields a post-trials
significance of 5.7 $\sigma$.   

To evaluate the extension and the position of the source, 
the sky-map was fitted to a simple symmetrical two-dimensional Gaussian
function, convolved with the instrument PSF (point spread function).  
The best-fit position lies at
$l=40.45^{\circ}\pm0.06_{stat}^{\circ}\pm0.06_{sys}^{\circ}$ and
$b=-0.80^{\circ}\pm0.05_{sta}^{\circ}\pm0.06_{sys}^{\circ}$, while the intrinsic
extension derived is $\sigma_{src}=(0.21^{\circ}+
0.07_{sta}^{\circ}-0.05_{sta}^{\circ}$). As the shape of the source
seems to depart from a symmetrical Gaussian, these values should be taken
as first approximations.

\begin{figure}[!b] 
\begin{center}
\includegraphics[width=0.51\textwidth]{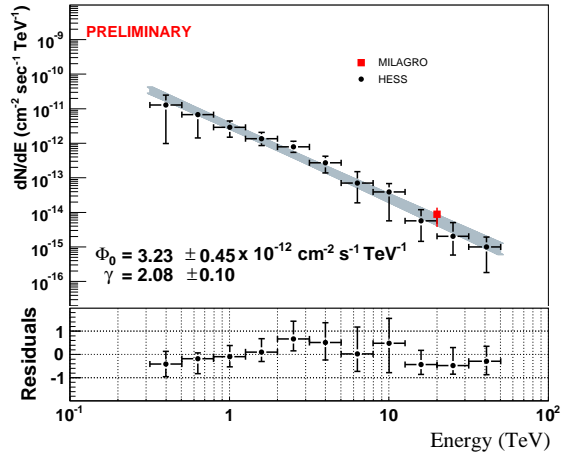}
\end{center}
\caption{Differential energy spectrum measured above 300~GeV for
HESS~J1908+063. The shaded area shows the 1 $\sigma$  
confidence region for the fit parameters. 
The differential flux of MGRO~J1908+06 at 20 TeV is shown in red. 
Fit residuals are given in the bottom panel.}  
\label{spectrum} 
\end{figure}

The differential energy spectrum was computed within an integration
radius of $0.5^{\circ}$ (corresponding to the FWHM of the source size
and shown as a white circle in Fig.~\ref{skymap}) 
centred on the best-fit position 
by means of a forward-folding maximum
likelihood fit \cite{CATSpectrum}. The spectrum is well fitted with a simple
power-law function (Fig.~\ref{spectrum}) with a hard photon index of
$2.08\pm0.10_{\rm stat}\pm 0.2_{\rm sys}$ 
and a differential flux at 1~TeV of ($3.23 \pm
0.45_{\rm stat} \pm 0.65_{\rm sys})\times 
10^{-12}$ cm$^{-2}$~s$^{-1}$. 
The integrated flux above 1~TeV corresponds to 14$\%$
of the Crab Nebula flux above that energy.

\section{Comparison with MGRO1908+06 \& Search for Counterparts}
\label{comparison}

\begin{figure}[ht]
\begin{center}
\includegraphics*[width=0.5\textwidth,angle=0,clip]{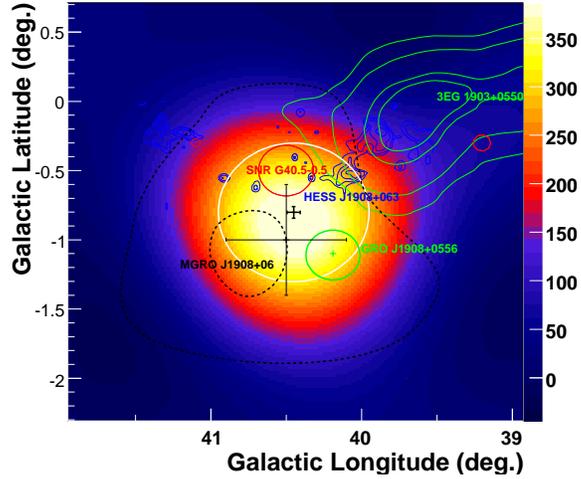}
\end{center}
\caption{ Multi-wavelength view of the 1.5$^{\circ}\times1.5^{\circ}$ field of
view around the position of HESS~J1908+063. The
    dotted black line shows the MILAGRO significance contours for 5 (inner) and
    8$\sigma$ (outer contour). The position of the EGRET GeV source
    GRO~J1908+0556 is marked with a green cross as well as the
    1$\sigma$ error in the position.  The 3EG 1903+0550 contours
    corresponding to 99, 95, 68 and 50$\%$ confidence levels are shown in
    green. The red circle marks the size and position of the radio-bright SNR
    G040.5-00.5. Contours in blue show the $^{13}$CO molecular cloud in the
    velocity range between (45,65) km/s.}
\label{skymapmwl}
\end{figure}

Fig.~\ref{skymapmwl} shows the $1.5^{\circ}\times1.5^{\circ}$ field of
view around the position of HESS~J1908+063 together with 
sources at other wavelengths including MGRO~J1908+06. The latter
source was discovered by the MILAGRO collaboration \cite{MILAGRO} 
after seven years of operation (2358 days of data) at the galactic
longitude and latitude of
$l=(40.4^{\circ}~\pm~0.1^{\circ}_{\rm stat}~\pm~0.3^{\circ}_{\rm sys}$) and  
$b=(-1.0^{\circ}~\pm~0.1^{\circ}_{\rm stat}~\pm 0.3^{\circ}_{\rm sys}$),
respectively.  The differential flux, at the median energy of 20~TeV, and
assuming a spectral index of -2.3, is at a level of  
(8.8$\pm$2.4$_{\rm stat}\pm$2.6$_{sys})\times
10^{-15}~{\rm TeV}^{-1}{\rm cm}^{-1}{\rm s}^{-1}$. MGRO~J1908+06 is reported to be 
both compatible with a point or extended source up to a diameter of
2.6$^{\circ}$. 

As clearly seen on Fig.~\ref{skymapmwl}, the positions of the two VHE sources
are fully compatible within errors. There is also a quite good
agreement between the differential flux at 
20~TeV of  MGRO~J1908+06 and the spectrum measured by HESS as shown on
Fig.~\ref{spectrum}. Given the larger integration radius of
$1.3^{\circ}$ for the MILAGRO source as compared to the $0.5^{\circ}$ radius
for HESS~J1908+063, the flux agreement implies the absence of any other
significant emission to the MILAGRO flux: the two sources can
consequently be identified to each other.

The better determination of the position of HESS~J1908+063 and the
measurement of its size and spectrum allow to search for counterparts
with stronger constraints.

At radio wavelengths, SNR~G40.5-0.5 \cite{Green} at
an estimated distance of 5.3~kpc overlaps with HESS~J1908+063. 
At EGRET energies, 3EG~J1903+0550, shown in green contours, lies close
to the SNR and has been suggested as possibly associated with it \cite{Sturner95}.
However G40.5-0.5 is not in exact coincidence with HESS~J1908+063
position and 3EG~J1903+0550 is only marginally overlapping with the
latter.  HEGRA observations of this region of
the sky \cite{HEGRAul} yielded an upper limit at 0.7~TeV at the SNR
position of 4.8$\%$ of the Crab Nebula flux.   
As this limit only applies for a point-like source it is not in
contradiction with the measurements reported here.

If the SNR is associated with the VHE source, the fact that the 22 
arc-min size of the shell is smaller than the FWHM of HESS~J1908+063
would contrast to previously discovered HESS sources identified
with shell-type VHE emitters, such as RX~J1713.7-3946
\cite{HESSG347} or RCW 86 reported at this conference
\cite{RCW86ICRC07}. The contribution of nearby unresolved sources or
interactions of accelerated cosmic rays with molecular matter in the vicinity of
the source could explain a larger size. However, for the latter
case, the position of the nearby $^{12}\rm CO$ cloud \cite{co} or
alternatively the $^{13}\rm CO$ contours (shown in blue on Fig.~\ref{skymapmwl}) do not favour
this scenario.

An analysis of the highest energy photons ($>$1~GeV) observed by EGRET
\cite{olaf97,lamb97} from this region shows a nearby and yet unidentified
source, GRO~J1908+0556/GEV J1907+0557. The positions of the two GeV
derivations are compatible within errors. GRO~J1908+0556, shown as a
green circle on Fig.~\ref{skymapmwl}, lies within a distance of less than
two times the EGRET 68\%  position measurement error to
HESS~J1908+063. A simple extrapolation of the H.E.S.S. spectrum to 
lower energies leads to a lower flux than that reported
for the EGRET source  ($6.33\times10^{-8}~{\rm cm}^{-1}{\rm s}^{-1}$). However
given the large PSF of EGRET even at GeV energies, other unresolved
sources can contribute to the flux measurement of GRO~J1908+0556. The
association of the HESS and MILAGRO sources to the GeV source is then
likely, although a coincidence by chance is not
excluded.

\section*{Summary}
In summary, a new source, HESS J1908+063 is reported above 300 GeV
at the level of 14$\%$ of the Crab Nebula flux and a post-trials
significance of 5.7~$\sigma$. The H.E.S.S.
source is extended, with a FWHM size of
0.5$^{\circ}$, and shows a hard spectrum with an index of 2.08$\pm$0.10. 
This detection confirms for the first time one of the low-latitude sources 
reported by the MILAGRO collaboration, MGRO~1908+062. 
A connection to the EGRET GeV source GRO~J1908+0556/GEV J1907+0557 at
lower energies remains possible. The association with SNR~G40.5-0.5 is not
excluded but the larger size of the TeV emission should then find an
explanation in terms of either contribution of unresolved sources or
interactions of ultra-relativistic particles with molecular matter in
the vicinity of the SNR. Deeper observations of this region with
Cherenkov telescopes and GLAST data would help the interpretation
of the detected VHE emission.

\section*{Acknowledgments}
The support of the Namibia authorities and of the University of Namibia
in facilitating the construction and operation of H.E.S.S. is gratefully
acknowledged, as is the support by the German Ministry for Education and
Research (BMBF), the Max Planck Society, the French Ministry for Research,
the CNRS-IN2P3 and the Astroparticle Interdisciplinary Programme of the
CNRS, the U.K. Particle Physics and Astronomy Research Council (PPARC),
the IPNP of the Charles University, the Polish Ministry of Science and
Higher Education, the South African Department of
Science and Technology and National Research Foundation, and by the
University of Namibia. We appreciate the excellent work of the technical
support staff in Berlin, Durham, Hamburg, Heidelberg, Palaiseau, Paris,
Saclay, and in Namibia in the construction and operation of the
equipment.

\bibliography{icrcHESS1908}

\begin{thebibliography}{10}

\bibitem{MILAGRO}
A.A. {Abdo {~ et al.}}
\newblock {\em astro-ph/0611691}, 2006.

\bibitem{HESSG347}
F.A. {Aharonian {~ et al. (H.E.S.S. Collaboration)}}.
\newblock {\em Nature}, 432:75--77, November 2004.

\bibitem{HESSScanII}
F.A. {Aharonian {~ et al. (H.E.S.S. Collaboration)}}.
\newblock {\em ApJ}, 636:777--797, January 2006.

\bibitem{HESSCrab}
F.A. {Aharonian {~ et al. (H.E.S.S. Collaboration)}}.
\newblock {\em A$\&$A}, 457:899--915, 2006.

\bibitem{HESSKooka}
F.A. {Aharonian {~ et al. (H.E.S.S. Collaboration)}}.
\newblock {\em A$\&$A}, 456:245--251, 2006.

\bibitem{HEGRAul}
F.A. {Aharonian {~ et al.(HEGRA Collaboration)}}.
\newblock {\em A$\&$A}, 439:635, 2005.

\bibitem{Green}
D.A. {Green {~ et al.}}
\newblock {\em A Catalogue of Galactic Supernova Remnants}, 2006.

\bibitem{HESSBack}
J.A.. {Hinton {~ et al.}}
\newblock In {\em Cherenkov 2005, Palaiseau}, 2005.

\bibitem{RCW86ICRC07}
S.~{Hoppe {\& M. Lemoine-Goumard for the H.E.S.S. Collaboration}}.
\newblock In {\em 30th ICRC, Merida, Mexico}, 2007.

\bibitem{HESSSurveyICRC07}
S.~{Hoppe {for the H.E.S.S. Collaboration}}.
\newblock In {\em 30th ICRC, Merida, Mexico}, 2007.

\bibitem{lamb97}
R.~C. {Lamb} and D.~J. {Macomb}.
\newblock {\em ApJ}, 488:872, 1997.

\bibitem{CATSpectrum}
F.~{Piron {~ et al.}}
\newblock {\em A\&A}, 374:895, 2001.

\bibitem{olaf97}
O.. {Reimer {~ et al.}}
\newblock In {\em 25th ICRC, Durban, South Africa}, page~97, 1997.

\bibitem{Sturner95}
S.~J. {Sturner} and C.~D. {Dermer}.
\newblock {\em A\&A}, 293:L17--L20, January 1995.

\bibitem{co}
J.~{Yang {~ et al.}}
\newblock {\em A\&A}, 6:210, 2006.

\end{thebibliography}
\bibliographystyle{plain}

\end{document}